\documentclass[10pt]{iopart} % iop

\usepackage{amssymb,amsfonts,mathrsfs,graphicx,url,hyperref}
\usepackage{cfr-lm}

\usepackage{microtype}

\begin{document}
\title{Event-chain Monte Carlo and the true self-avoiding walk}
\author{A.~C.~Maggs}
%\affiliation{CNRS UMR 7083, ESPCI Paris, Université PSL,  10 rue Vauquelin,
%75005 Paris, France.} %aps
\address{CNRS UMR 7083, ESPCI Paris, universit\'e PSL, 10 rue Vauquelin, 75005
  Paris, France.}
% \email{anthony.maggs@espci.fr} %aps
 \ead{anthony.maggs@espci.fr}
\begin{abstract}
  We study the large-scale dynamics of event chain Monte Carlo algorithms in one
  dimension, and their relation to the true self-avoiding walk. In particular,
  we study the influence of stress, and different forms of interaction on the
  equilibration and sampling properties of algorithms with global balance, but
  no local balance.
\end{abstract}
%\maketitle
\section*{Introduction}

Irreversible Monte Carlo algorithms which rely on global balance~\cite{Wilson,
  Manon}, while not obeying the more restrictive criterion of detailed balance
have had remarkable successes in the study of hard numerical problems such as
the nature of ordered phases of two-dimensional fluids~\cite{bernard}. A
fundamental question is the origin of the speed-ups compared to classical
algorithms, such as molecular dynamics and reversible Monte
Carlo~\cite{Kapfer}. It is clear that such irreversible algorithms are rather
unusual for dynamical systems sampling the equilibrium Boltzmann
distribution. They display large-scale coherent flows, in a manner that recalls
active matter. The collision rules are however finely tuned so that these flows
do not disturb the equilibrium state. Indeed, it is because of these large-scale flows
that irreversible methods can explore configuration space more efficiently than
diffusive algorithms such as Monte Carlo.

In a recent paper~\cite{nonrev} we have demonstrated numerically that
event-chain Monte Carlo (ECMC) simulation algorithms applied to a harmonic chain
are a realization of the so-called ``true'' self-avoiding walk~\cite{Amit,
  PELITI1984225, rammal}. Remarkably the true-self avoiding walk has been
studied by the mathematical community and solved analytically in
one-dimension~\cite{Toth, DUMAZ20131454,Wendelin}, opening the possibility of better
understanding the large-scale dynamic behaviour of irreversible
algorithms. The detailed numerical comparisons that we made 
considered only a harmonic chain in order to remain as close to the exact
mathematical results as possible. In this paper we consider a larger range of
physical systems and potentials, in order to study the universality of the findings.

The mathematical literature on the true self-avoiding walk emphasizes the
importance of two distribution functions. The first $\rho_1(t,x)$ describes the
distribution of end-to-end separations, $x$, in a polymer growth problem. When
studying irreversible Monte Carlo methods $x$ maps onto the final position (or
rather the label) of the mobile particle at the end of a simulation in which the
total displacement of all particles is equal to $t$.  The second distribution
$\rho_2(t,h) $ gives the distribution in the number of previous visits, $h$ to a
position for the polymer problem. In the context of Monte Carlo simulation it
corresponds the distribution in the number of updates to a particle. For the
growing polymer, it was demonstrated that in the one-dimensional continuum
limit, 
\begin{eqnarray}
  \rho_1(t,x) &=& t^{-2/3} \nu_1( |x| t^{-2/3}) \label{eq:nu1}\\
  \rho_2(t,h) &=& t^{-1/3} \nu_2 (h t^{-1/3} )\label{eq:nu2}
\end{eqnarray}
where the scaling functions $\nu_1$ and $\nu_2$ are known
explicitly~\cite{DUMAZ20131454}. %In our presentation we will also work with
% the scaled variables $\tilde x = x t^{-2/3}$, $\tilde h=h t^{-1/3}$
They are plotted as solid, red lines in Fig.~\ref{fig:cold1}.
These scaling functions were confirmed to apply to the ECMC simulation of  tension-free harmonic
chain.  However, in the applications of ECMC with non-trivial potentials this
scaling form can not appear: Eq.~(\ref{eq:nu1}) is symmetric under
$x \rightarrow -x$, so that the distribution does not drift with time. In ECMC
the position of the mobile (active) particle drifts with a constant
speed~\cite{Manon}, equal to the thermodynamic pressure, so that $\rho_1(t,x)$
can not remain symmetric around the origin, $x=0$.

The drift, linked to the pressure, can be cancelled by modifying the microscopic
interactions on a chain with the addition of an extra linear potential (factor
field)~\cite{LeiFF}. This modified potential is built in such a way that it does
not change the thermodynamic properties of the simulated system, only the
dynamics through the ECMC update rules. We thus modify the potential between two particles so that
\begin{equation}
  V(r) \rightarrow V(r) + p\, r \label{eq:ff}
\end{equation}
Remarkably this modification was found to
give rise to further acceleration in the sampling of the underlying physical
system.  For the optimal choice of the amplitude of the linear potential (factor
field) the dynamic exponent $z$, describing the relaxation of long-wavelength
density fluctuations, takes on the low value of $z=1/2$, rather than the
exponents $z=1$ and $z=2$ characteristic of molecular dynamics and Monte Carlo
methods, and the value $z=1$ characteristic of the historic ECMC method~\cite{LeiFF}. This
acceleration is directly linked to the exponent $2/3$, appearing in
eq.~(\ref{eq:nu1}). Indeed, the super-diffusive propagation displayed in
eq.~(\ref{eq:nu1}): $|x| \sim t^{2/3}$ is just linked via scaling to the dynamic
exponent: $t\sim |x|^{(1+z)}$.  Let us also note that a lifted
TASEP~\cite{essler2023lifted} (Totally Asymmetric Simple Exclusion Process), is
believed to have similar scaling properties to ECMC with factor fields and can
be studied with the help of Bethe methods. We concluded that there exists a
dynamic universality class~\cite{nonrev} for one-dimensional systems which is distinct from
that which has been widely studied for the KPZ system~\cite{kpz}.

The question explored in the present paper is the degree to which this
universality class is stable to perturbations in the dynamics coming from
modifications in the properties of the underlying physical system.  We consider
interactions which are more complicated than the harmonic springs
of~\cite{nonrev}, but also a broader range of simulation protocols, such as hot
and cold starts as well as quenches in order to see how stable the observed
behaviour is. We also study the effect of stress in modifying the distribution
of eq.~(\ref{eq:nu1}).

\section*{Implementation of irreversible algorithms}
For the implementation of the event chain Monte Carlo  we follow the
presentation of~\cite{Faulkner}. The total potential energy of a physical system
is broken up into a sum of independent factors. We take as the factors the pair
energy of particles in a chain. Motion of a single  particle leads to
changes in the pair energies, which is then used in an individualized Monte
Carlo criterion (the factorized Metropolis algorithm). This individualized
criterion allows one to choose a collision partner among all the changing
factors, and to continue the motion with a new mobile particle without
generating a Monte Carlo rejection. The unique mobile particle at any moment is
called ``active''. If the pair potential between two particles is $V(r)$ finding the
candidate moment of collision, when motion is transferred to a new particle, requires
solving an equation of the form
\begin{equation}
  \Delta  V^+(r) =-T \ln{(\mathrm{rand})} \label{eq:energy}
\end{equation}
$\mathrm{rand}$ is uniformly distributed on $(0,1)$. $ T=1/\beta$ is the
temperature.  $V^+$ is constructed from a clipped derivative of $V$:
\begin{equation}
  \frac{dV^+(r)} {dr} = \max \left (0,  \frac{dV(r)} {dr}  \right)
\end{equation}
One then compares all the candidate collisions and takes the very first; for a
chain with nearest-neighbour interactions this requires the comparison between
two candidate events.
The implementation is rendered more complicated, because in general we are
interested in the addition of an extra linear potential (factor
field)~\cite{LeiFF} to the bare potential $V(r)$, eq.~(\ref{eq:ff}).

In this paper we consider three cases. Firstly harmonic chains, where it is
possible to find the solution to eq.~(\ref{eq:energy}) with elementary functions. Secondly, the case of
exponentially repelling particles for which the Lambert-$W$
function~\cite{lambertw} allows a direct solution to the energy equation.
Finally, Lennard-Jones interacting particles, for which we use iterative root
solvers for eq.~(\ref{eq:energy}). The Lennard-Jones system is studied at a low temperature where the
system breaks into small clusters so that the dynamics is  highly
heterogeneous.

These different models allow us to explore the universality of the distribution
function $\rho_1(t,x)$ as well as test how to implement factor field
acceleration in the case that interactions go beyond nearest neighbours. We
explore interactions with first and second-nearest neighbours, and show how to
implement factor field acceleration in this more general system.

\section*{Harmonic chains}

We firstly extend our study of the harmonic chain~\cite{nonrev}. Our
previous work only considered the dynamics of a system prepared in its ground
state (cold starts), in a state of zero tension. We firstly study the effects of
temperature jumps on the dynamics.

We take as the energy
\begin{equation}
  E= \frac{k}{2}\sum_{i=1}^N {(y_i -y_{i+1})}^2 \label{eq:harmonic}
\end{equation}
with $y_i\equiv y_{N+i}$. $k$ is a spring constant. The quadratic energy
function of a harmonic chain with nearest-neighbour interactions can be written
in terms of a sparse (mostly zero-filled) matrix, $E={y^T}My/2$. This matrix
$M$, admits a Cholesky factorization, $M=R^T R$, where $R$ is an upper
triangular matrix, again with sparse structure.  An equilibrated sample is then
found by solving the equation $R y=\xi$, where $y$ is the configuration and
$\xi$ Gaussian distributed, independent random numbers. In order to remove the
zero-mode that occurs in $M$, we attach the last site of the chain to a
reference position, with an extra spring, during the factorization step but not
during the ECMC simulation. This procedure allow us to
easily generate equilibrated systems, the algorithmic complexity scaling
linearly with the system size.

\begin{figure}
\center{\includegraphics[width=.6\columnwidth]{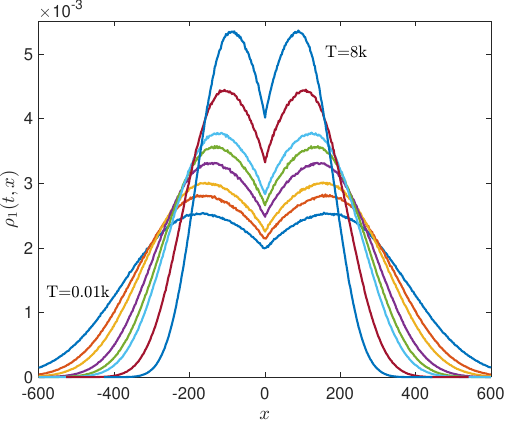}}
  \caption{A zero tension harmonic chain is prepared at temperature $T/k=1$, ECMC
    simulations are then performed at:
    $T/k= [0.01, 0.25, 0.5, 1, 1.5, 2, 4, 8]$.  Evolution, $\rho_1(t,x)$,
    eq.~(\ref{eq:nu1}) for event chain lengths of length $t=2048$.
    %\protect\url{imac::Avoid_run/Gaussian/go_T.sh+superT.m}.
    Higher
    temperatures spread more slowly giving narrower, higher curves.  All curves
    superpose to within line width on scaling to a standard width. Data from
    $2^{25}$ configurations for each curve. %As a comparison,
  }\label{fig:quenching}
\end{figure}
 
\begin{figure}
\center{\includegraphics[width=.6\columnwidth]{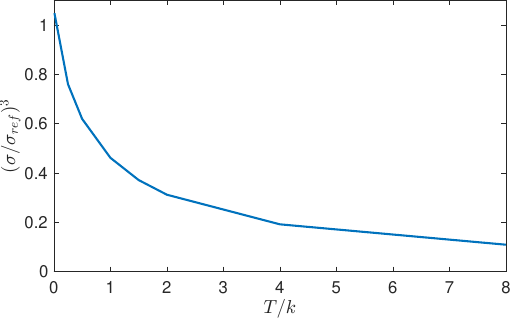}}
  \caption{Evolution of the width, $\sigma$, of distribution $\rho_1(t,x)$,
    eq.~(\ref{eq:nu1}) from the data in Fig.~\ref{fig:quenching}. We normalized
    to width, $\sigma_{ref}$ in a system prepared at zero temperature and then
    simulated at $T/k=1$.
    %\protect\url{imac:Avoid_run/Gaussian/go_T.sh+superT.m}.  %As a comparison,
  }\label{fig:T}
\end{figure}

\begin{figure}
\center{\includegraphics[width=.6\columnwidth]{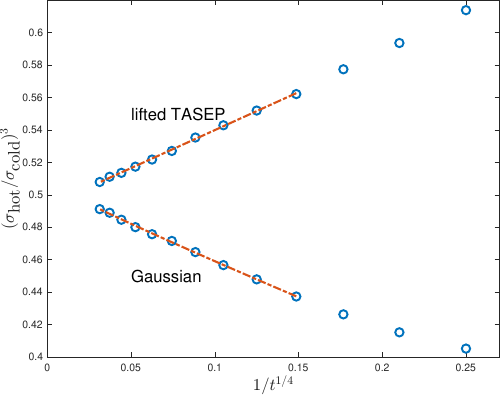}}
  \caption{Ratio of widths $\sigma_{\textrm{hot}}$ and $\sigma_{\textrm{cold}}$
    for hot and cold starts, as a function of  $t^{-1/4}$. The
    long-time limits extrapolate to 
    ${(\sigma_{\textrm{hot}}/\sigma_{\textrm{cold}})}^3$ close to $1/2$. Times
    from $t=2^8$ to $t=2^{20}$. Red dashed line fit to eq.~(\ref{eq:extrap}).
    %\protect\url{Avoid_run/Gaussian/Ratio/res.m,+whistler+larix+sargent}.  %As a comparison,
  }\label{fig:hotcold}
\end{figure}

We generate initial configurations equilibrated at $T/k=1$, then run the
algorithm at different temperatures, if the simulation temperature $T/k<1$, this
imposes a quench on the physical system. If $T/k>1$ it is a sudden heating. We
measure the distribution function $\rho_1(t,x)$ after a constant time $t$
(corresponding to the total displacement of particles during a run), and plot
the results in Fig.~\ref{fig:quenching}. For all changes in
temperature the evolution is remarkably similar, the distribution displays the
double peak structure familiar from the analytic solution of the true
self-avoiding walk eq.~(\ref{eq:nu1}). When we rescale these curves they
superpose to within the displayed line width; the scaling function is
insensitive to the surrounding environment. In Fig.~\ref{fig:T} we plot the
width of the distribution (root-mean-squared deviation), $\sigma^2=\langle x^2 \rangle$, as a function
of the final temperature. In this figure we scale the width by the width of a
zero-temperature starting configuration propagating at $T/k=1$, the protocol
that was studied in our previous publication~\cite{nonrev}.  As temperature
increases the distribution narrows.

Exact calculations in~\cite{DUMAZ20131454} on the true-self avoiding walk
indicate  that if we compare the propagation of a cold start, and a hot
start the ratio of  widths
${(\sigma_{\textrm{hot}}/\sigma_{\textrm{cold}})}^3=1/2$ This corresponds to the
point $(1,1/2)$ in Fig.~\ref{fig:T}. In Fig.~\ref{fig:hotcold} we study
${(\sigma_{\textrm{hot}}/\sigma_{\textrm{cold}})}^3$ as a function of simulation
time $t$ in order to better evaluate the infinite time limit of the
ratio. Empirically we find a near-linear extrapolation if we plot the data as a
function of $1/t^{1/4}$. We performed a non-linear fit of our data for large $t$
to the form
\begin{equation}
  {(\sigma_{\textrm{hot}} / \sigma_{\textrm{cold}} )}  ^3 = \alpha_1 + \frac{\alpha_2}{t^{\alpha_3}} \label{eq:extrap}
\end{equation}
using the Matlab function ``nlinfit''. The result is shown as a dashed line in
the~Fig~\ref{fig:hotcold}.  We find
\( (\alpha_1, \alpha_2, \alpha_3 ) = (0.507, -0.414, 0.233) \).  This very slow
extrapolation with time prevents us from finding a high precision result, but
the results for this ratio for ECMC do seem compatible with the mathematical
result for the true self-avoiding walk.

We performed analogous simulations for the lifted TASEP at half filling. We
compare starting configurations of either an ordered state  (``cold''
start) or configurations drawn from the equilibrium distribution (``hot''
start). The data displays a very similar scaling as a function of simulation time
giving, \( (\alpha_1, \alpha_2, \alpha_3 ) = (0.493, 0.470, 0.251) \)

Both extrapolations are rather remarkable, since the microscopic formulation
of ECMC on harmonic chains, or lifted TASEP is rather different from the
formulation of the  true
self-avoiding walk. We conclude that there is universality in scaling ratios
for ECMC, as well as the scaling function itself.

\subsection*{Stress}
Let us now consider the simulation of a harmonic system with  strain
with the energy
\begin{equation}
  E= \frac{k}{2}\sum_{i=1}^N {(y_{i+1} -y_{i} -l_0)}^2  \label{eq:l0}
\end{equation}
The ground state of this energy is still $y_i=\mathrm{const}$. This corresponds
to stressing each individual spring, by a constant value $p=k l_0$. From the
results of~\cite{Manon, LeiFF} this must lead to a change in the large scale
dynamics of the algorithm since the stress and mean displacement of the active
(mobile) particle are linked by \( p = \langle x \rangle/t\). The average
position of the distribution \(\rho_1(t,x)\) thus displaces with increasing
$l_0$

\begin{figure}
\center{\includegraphics[width=.6\columnwidth]{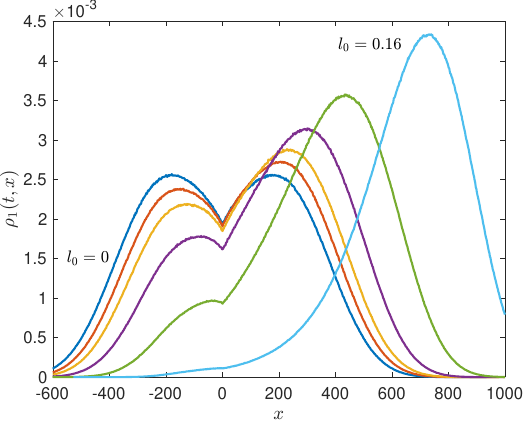}}
  \caption{Evolution of distribution, $\rho_1(t,x)$, eq.~(\ref{eq:nu1}) with
    stress eq.~(\ref{eq:l0}) $l_0=[0, 0.01, 0.02, 0.04, 0.08, 0.16]$. $N=8192$,
    path length $t=2048$.
    $T/k=1$.
    %\protect\url{Avoid_run/Gaussian/go_l0.sh+super2.m}.
    Data from
    $2^{26}$ simulation for each curve.}\label{fig:l0}
\end{figure}

We perform simulations starting with a ground state configuration $y_i=0$ (cold
start). We see Fig.~\ref{fig:l0} that imposing a stress leads to a drift of the
configuration to the right; we checked that the drift speed is proportional to
$l_0$. Rather remarkably, for small to moderate values of $l_0$ the singularity
at the origin in the curve $\rho_1(t,x)$, remains even as the centre of mass of
the curve transports
to the right.
% $l_0$. %It is only for larger $l_0$ that we generate a packet
% of sampled sites that eliminates this singularity.

% \subsection*{Gaussian chain, second-nearest neighbor interactions}

\section*{General potentials}
\subsection*{One dimensional statistical mechanics}
The statistical mechanics of a system of particles with nearest-neighbour
interactions can be solved analytically in the isobaric
ensemble~\cite{feynman}. In particular the distribution of particle separations
follows the distribution law \( e^{-\beta(V(r)+p\, r)} \), where $V(r)$ is the
inter-particle potential for separation $r$, and $p$ the thermodynamic
pressure. From this distribution we find the average separation between
particles
\begin{equation}
  \Delta(p) = \langle r \rangle_p=  \frac{\int r\,  e^{-\beta(V(r)+p\, r)}\, dr} {\int   e^{-\beta(V(r)+p\, r)}\, dr} \label{eq:dist}
\end{equation}
With this statistical weight we also find the useful relations:
\begin{equation}
    \langle r dV/dr \rangle_p +p \Delta(p) =T \label{eq:virial}
\end{equation}
\begin{equation}
    T \rho_c- \langle dV/dr \rangle_p =p \label{eq:force}
   \end{equation}
Eq.~(\ref{eq:virial}) is the usual virial
equation~\cite{period}, that we use to verify the stress state of our chains. In
 eq.~(\ref{eq:force}) the contact density, $\rho_c = e^{-\beta V(0)}/z_1$, with $z_1$ the single bond
partition function. When the potential diverges at the origin, as is the case
for the Lennard-Jones potential, $\rho_c=0$, and we have an easy, independent measure
of the thermodynamic pressure.

In our simulations we use both cold and hot starts. For the cold start we
initialize with a uniformly spaced system, where we calculate, by numerical
integration of eq.~(\ref{eq:dist}), the mean separation as a function of
pressure: and place particles uniformly, with a total system length of
$L=N \Delta(p)$. We then use the imposed $p$ as the value of the factor field in
our simulations.

For the hot start (where a sample is pre-equilibrated) before each simulation we
generate a library of $10^8$ samples according to the distribution of
eq.~(\ref{eq:dist}) using Chebyshev interpolation to implement  inverse
transform sampling~\cite{chebfun}. From this
library of separations we generate large numbers of initial states needed to
calculate the distributions $\rho_1(t,x)$ and $\rho_2(t,x)$ by randomly drawing
a series of $N$ values of $r$ from this library. We expect that differences from
the isobaric and constant length ensembles are small for the large system sizes
that we simulate.

\subsection*{Longer range interactions}

There is no closed form expression for the properties of one-dimensional chains
with interactions beyond first-nearest neighbour. It is not possible
to generate pre-equilibrated samples as above. We can still, however, use the virial
theorem to express the pressure in terms of the potential. We specialize to the
case of first and second-nearest neighbour interactions. We have:
\begin{equation}
p  L  = NT + \sum_{i<j} r_{ij} f_{ij}
\end{equation}
where $r_{ij} = r_i-r_j$, with the nearest image convention and $f_{ij}$ is the
force on $i$ from $j$. Let us now introduce factor fields $p_1$ and $p_2$,
between first and second-nearest neighbours, so that for instance the nearest
neighbour interaction becomes $V_1(r) +p_1\, r$. We find from the virial theorem
with the modified potentials:
\begin{equation}
  (L/N) p = T + \langle r^{(1)} f_1  + r^{(2)} f_2 \rangle - (L/N) (p_1 + 2p_2) 
\end{equation}
where $f_i(r) = - dV_i(r)/dr$, and $r^{(i)}$ the corresponding separation vector.
We anticipate that the zero drift criterion for the distribution $\rho_1(t,x)$,
then corresponds to $p=p_1+2p_2$. While the zero-drift criterion gives a
unique criterion for the factor field with nearest neighbour interactions, this
is no-longer the case when longer-ranged interactions are present. We 
explore numerically the dynamics to determine the efficiency of algorithms with
different choices of $p_1$ and $p_2$. In particular, we make a detailed study
with repulsive exponential interactions.

We generalize the result of eq.~(\ref{eq:force}) using the methods
of~\cite{historic}.  We consider a chain of $N$ particles with first and
second-nearest neighbour interactions $V_1$, $V_2$. We then shorten the chain by
$\epsilon$. Clearly $\beta p= -\partial \ln{Z}/ \partial \epsilon$. We implement this
shortening by removing a slice in configuration space between $0$ and
$\epsilon$. We hold a single $i=1$ particle at $r=0$, to fix the centre of mass
of the chain. Then, for instance the integral over the particle $i=2$ becomes,
\[
  \int_\epsilon^{r_3} dr_2\, \exp{(-\beta V(r_2-r_1-\epsilon))}
\]
With a similar modification for the interaction with the second-nearest
neighbours. Keeping variations to order $\epsilon$ we find
\[
  p = T \rho_c - \left \langle \frac {\partial V_1} {\partial r} \right \rangle
  - 2 \left \langle \frac {\partial V_2} {\partial r} \right \rangle
\]
The factor of two is due to the fact that shortening a single nearest neighbour
separation modifies the distance between two particles at second-nearest
neighbour. The erm in $\rho_c$ comes from the disappearance of 
configurations at contact between the particles $i=1$ and $i=2$.

\subsection* {Exponential interactions}

\begin{figure}
\center{\includegraphics[width=.6\columnwidth]{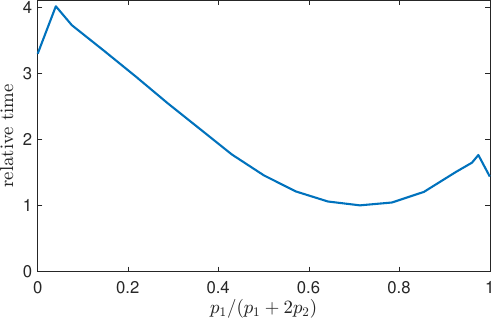}}
  \caption{Autocorrelation times of chain with first and second-nearest neighbour
    exponential potentials, on the line $p=p_1+2p_2$. Simulation time, $t=2^{29}$, $N=8096$.
    % \protect\url{imac:github/Avoid_dev/lambert/virial/p.pdf+wilson.m}
}\label{fig:autocorr}
\end{figure}

We consider the case of repulsive interactions between particles of the form
$V(r)=a \exp(-|r|/\ell)$ and systems with first and second neighbour
interactions. To implement the event-chain algorithm, with a factor field we
need to find the solution, $r$, to the equation
\begin{equation}
  a \exp(-r/\ell) + p r =\Delta E \label{eq:lambertfactor}
\end{equation}
$\Delta E$, is calculated from the difference between $V$ and $V^+$, together
with the thermal activation.  Solutions to eq.~(\ref{eq:lambertfactor}) are
found from the Lambert-$W$ function~\cite{boost}, defined as solutions to the
equation
\begin{equation}
  W e^W=z
\end{equation}

We study two questions with this interaction, firstly the universality of the
functions $\rho_1(t,x)$ and $\rho_2(t,x)$ when non-longer working with harmonic
springs and secondly how the factor-field method generalizes to systems with
longer-range interactions.  Let us emphasize that the factor-field method, as
previously studied, introduces a linear potential between nearest-neighbour
interacting particles, with an amplitude which is equal to the thermodynamic
pressure. It is at this unique point that the method is the most efficient. With
more general interactions there is potentially a line of solutions
$p_1+2 p_2 = p$, where $p_i$ is the factor field with the $i'th$ neighbour and
$p$ is again the thermodynamic pressure. In order to generate an efficient
algorithm are their further constraints on the individual amplitudes $p_i$?

We implemented the simulations with $a=20$, for nearest neighbour interactions
and $a=40$, for second-nearest interactions, with $\ell=2$, $T=0.3$,
$L=4.69703 N$, $N=8192$. Preliminary simulations without
factor fields are used to estimate the thermodynamic pressure from the virial
theorem.  We equilibrate the system, and then calculate the
autocorrelation time of the lowest Fourier mode of the density.  %Study
%the distribution  functions $\rho_1(t,x)$ and $\rho_2(t,x)$ shows 
%that they are indeed compatible with the analytic forms
%eqs.~(\ref{eq:nu1},~\ref{eq:nu2})
 We then perform  simulations 
along the line $p_1+2 p_2 = p$. The dynamics are characterized by the
universal scaling forms, eq.~(\ref{eq:nu1},~\ref{eq:nu2});
this remains true even for the extreme cases $p_1=0$ or $p_2=0$.

When measured in terms of the Monte Carlo time $t$ (equal to the total
displacement of particles) the code efficiency is almost
independent of the exact mix between $p_1$ and $p_2$, however extreme choices 
gives rise to small steps before generating a collision event, so that the
clock time of a simulation is sensitive to the exact mix of fields. %; more collisions are then
%generated in the time $t$ requiring more calculation.
In Fig.~\ref{fig:autocorr} we plot a relative clock time (compared to the best
mix of fields) as a function of $p_1/(p_1+2p_2)$, demonstrating a relatively
broad minimum for the autocorrelation time, in units of wall clock time. Note
the two end points for $p_1=0$, and $p_2=0$ deviate from the main curve, since
fewer evaluations of the Lambert-W function are required. We conclude that the
universal scaling form remains stable with interactions which go beyond nearest
neighbour, and that the efficiency of ECMC depends only weakly on the exact
field values, if the factor fields are tuned correctly to the
thermodynamic pressure.

\subsection*{Lennard-Jones systems at low densities}

\begin{figure*}
  \includegraphics[width=0.95\textwidth]{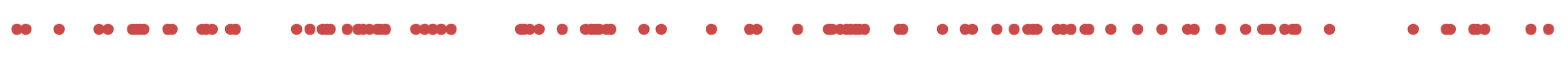}
  \caption{Equilibrated Lennard-Jones configuration at a low density, showing
    well separated clusters. $\beta \epsilon=2$, $\beta p=0.03$, $N=96$ particles,
    mean separation.
%    \protect\url{imac:/Users/tony2/github/Avoid_run/lj}.
    $\Delta=4.007$.}\label{fig:config}
\end{figure*}

We implemented factor-field simulation of particles with nearest-neighbour
Lennard-Jones interactions. A previous paper~\cite{LeiFF} reported the properties of
this system at high densities, such that the average separation between
particles is smaller than the position of the potential minimum. We here study
the properties of a system in which $\beta p=0.03$, $\Delta=4.007$, and for
which $\epsilon \beta=2$, where $\epsilon$ is the well depth. Length units are
set by the Lennard-Jones potential which crosses $E=0$ for unit separation. As
seen in Fig.~\ref{fig:config} for these parameters the system breaks up into a
series of isolated clusters. The ECMC algorithm must equilibrate the internal
structure of the clusters, as well as ``jump'' the activity between
clusters.  The question is then whether the large scale dynamics of this
heterogeneous system still converges to the universal forms
of~\cite{DUMAZ20131454, nonrev}, eq.~(\ref{eq:nu1},~\ref{eq:nu2})

\begin{figure}
\center{\includegraphics[width=.6\columnwidth]{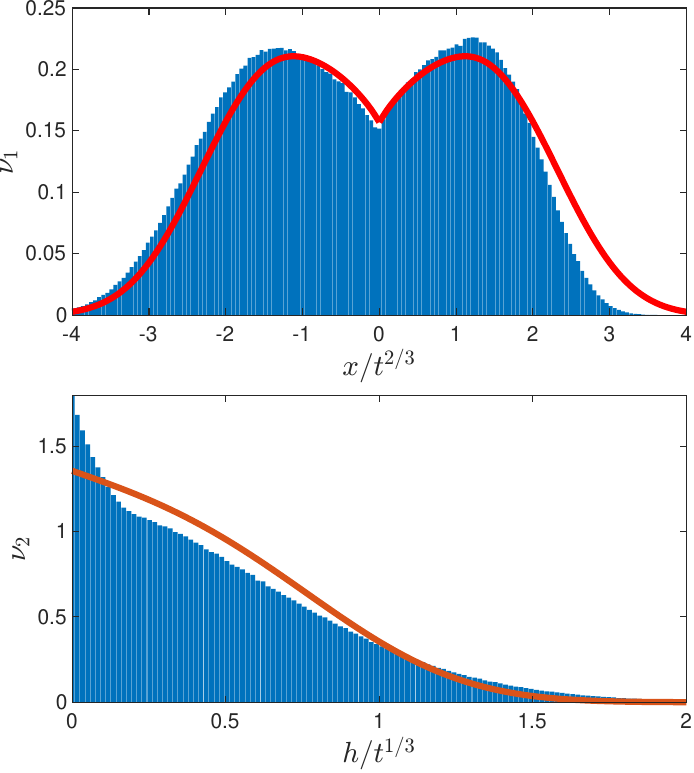}}
 \caption{Plots of the scaling functions $\nu_1$ and $\nu_2 $,
    eqs.~(\ref{eq:nu1},~\ref{eq:nu2}) for a cold start with simulation time
    $t=1024$, parameters correspond to Fig.~{\ref{fig:config}}. For short times
    $\rho_1(t,x)$ has a clear back-forwards asymmetry, which only slowly decays
    for longer simulation times, Fig.~\ref{fig:skew}. $\rho_2(t,h)$ has a
    pronounced peak for small $h$. Red, solid lines the analytic solution to the
    true self-avoiding walk eq.~(\ref{eq:nu1},~\ref{eq:nu2}) Blue histogram:
    binned data from simulation.
    % \protect\url{sargent:/home/tony/lj/cold_run/ColdRerun.1024/nu_tile.pdf+nu_tile.m}.
}\label{fig:cold1}
\end{figure}

Implementation of the factor-field algorithm for the Lennard-Jones potential
requires solving equations of the form
\begin{equation}
  \frac{1}{r^{12}} - \frac{1}{r^6} + p r = \Delta E \label{eq:ljfactor}
\end{equation}
for $r$. % We note that in these units the well depth $\epsilon$ of the
% Lennard-Jones potential is $\epsilon=1/4$. This occurs at a separation
% $r_0= 2^{1/6}\approx 1.12$
We proceed by using an iterative solver, based on Halley iteration~\cite{boost},
a higher order generalization of Newton-Raphson. From an initial guess of the
root  it requires typically four iterations to fully converge
the solution of eq.~(\ref{eq:ljfactor}) to machine precision.

We find that the ``hot'', pre-equilibrated samples generate a symmetric
function $\rho_1(t,x)$ in agreement with eq~(\ref{eq:nu1}). However, the cold
samples generate a small back-forward asymmetry for $\rho_1(t,x)$, Fig.~\ref{fig:cold1}. The
distribution of $\rho_2(t,x)$ differs considerably from the analytic form
Fig.~\ref{fig:cold1}, with a strong peak for small $h$. This implies that if one
stops the simulation one has visited the final visited site much more rarely than
would be expected from the statistics of the true self-avoiding walk.

\begin{figure}
\center{\includegraphics[width=.6\columnwidth]{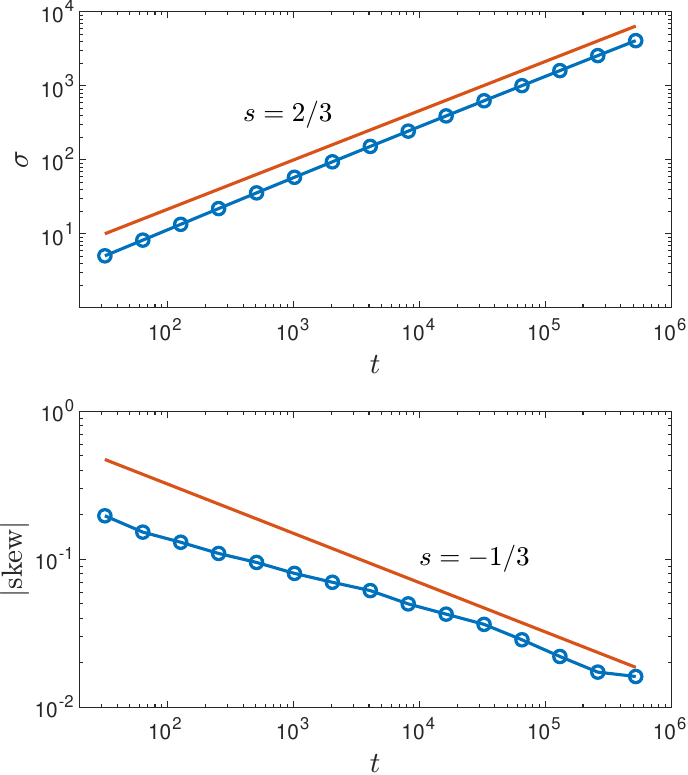}}
  \caption{Top: evolution of the width, $\sigma$, of the distribution of
    $\rho_1$ as a function of simulation time $t$.  Red line as a guide to the
    eye with exponent $s=2/3$. Bottom, evolution of the skew with simulation
    time $t$. Red line exponent $s=-1/3$ guide to eye. System corresponds to
    Fig.~\ref{fig:config}.
    % \protect\url{sargent:/home/tony/lj/cold_run/skew.pdf+mom.m}.
    Data from
    $2^{22}$ configurations per point.}\label{fig:skew}
\end{figure}

We study the skewness of the distribution $\rho_1(t,x)$, Fig.~\ref{fig:skew}
bottom, as a function of simulation time $t$, where the skewness is defined as a
normalized third moment:
\begin{equation}
  \mathrm{skew} =  \langle     {(x - \langle x \rangle )}^3 \rangle/\sigma^3 \label{eq:skew}
\end{equation}
We find that eq.~(\ref{eq:skew}) decays only very slowly with $t$, and possibly
at the largest times decays as $\mathrm{skew} \sim t^{-1/3}$, which is similar
to the asymmetry found in the lifted TASEP model~\cite{essler2023lifted}.
Despite the slow decay of the skewness of the distribution to zero we find that
the width of the distribution, $\sigma$, Fig.~\ref{fig:skew} top, fits the
exponent $2/3$ from eq.~(\ref{eq:nu1}) even for short times. Despite showing
very slow convergence in time, we conclude that even this heterogeneous system
does display the universal true self-avoiding form.

\section*{Mixing times}

Until now, we have used autocorrelation times as a measure of efficiency of a
simulation algorithm. However, when starting from an arbitrary configuration one
should also have some idea of mixing times, which bound the time for samples
generated by the simulation to be close to equilibrium, after starting in an
arbitrary state. Certainly in physical applications there are many examples
involving nucleation, where this time can be much larger than autocorrelation
times. For one-dimensional models with factor-field accelerated ECMC we
showed~\cite{LeiFF} that a system of hard rods, which has an autocorrelation
time for density fluctuations scaling as $N^{3/2}$ has an asymptotically slower
mixing time scaling as $N^2$. This is for a configuration where all particles
start in contact, and relaxation of the configuration only occurs by loss of
particles from the end of a dense starting state. Can one make similar
statements for the simulation of models such as the chains considered in this
paper? For simplicity, we will only consider configurations of the harmonic
chain, to avoid considering configurations such as that displayed in
Fig.~\ref{fig:config} where the use of very deep Lennard-Jones potentials and
very low densities must  lead to slow coarsening dynamics.

We proceed by studying trial configurations, that are far from equilibrium, and
make analytic arguments which we confirmed with numerical studies. We start with
an analogue of the dense configurations which have the largest mixing times for
hard rods:
\begin{eqnarray}
%\begin{align}
  y_i=& a\quad\quad 1\le i\le \lfloor N/2 \rfloor \label{eq:step}\\
  =&-a\quad\quad \textrm{otherwise}\nonumber
     %  \end{align}
     \end{eqnarray}
$\lfloor \, \rfloor$ denotes the integer part and $a>0$ is an amplitude.  
On starting the position of the active particle at random it will take 
a time $O(N^{3/2})$ to find the region where $y_i<0$. For $a^2k \gg T$ the
active particle then remains confined to regions with $y_i<0$ while making steps
of amplitude $O(\sqrt{T/k})$. Thus, it takes a time $O(aN\sqrt{k/T})$ to erase
the step in the initial configuration with ECMC. If $ a^2\gg N {T/k}$ then this
time is longer than the autocorrelation time. We can easily build configurations
for which the mixing time is unbounded. The case $a^2=N {T/k}$ is
interesting. It corresponds to injecting energy $O(N T)$ into the chain;
comparable to the thermal energy at equilibrium. 
This motivates two
questions.
\begin{itemize}
\item (1) Are there configurations with energy $\gg N T$ which nevertheless
  mix rapidly?
\item (2) Are there configurations with an energy budget $O(NT)$ that mix more
  slowly than  eq.~(\ref{eq:step})?
\end{itemize}

\begin{figure} 
\center{\includegraphics[width=.5\columnwidth]{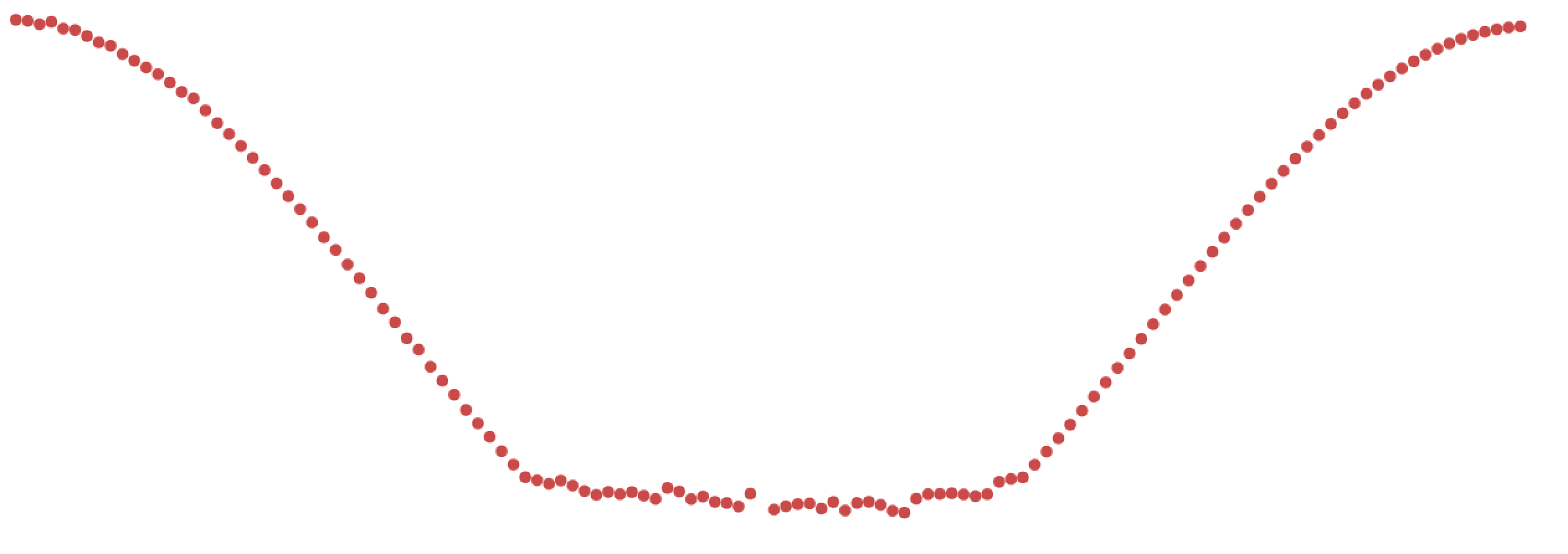}}
  \caption{Initial configuration is chosen as eq.~(\ref{eq:cos}). The algorithm
    has driven the active particle to the central part of the configuration,
    which filled in with a local state which is close to
    equilibrium. When the ECMC has erased the whole initial state
    we find a bound on the mixing time.}\label{fig:cosfig}
\end{figure}

To answer case (1) we consider the configuration
\begin{eqnarray}
%\begin{align}
  y_i &=a\quad \quad i \textrm{ odd} \\
  y_i &=-a\quad \quad i \textrm{ even} \nonumber
        %  \end{align}
        \end{eqnarray}
The energy is unbounded for large amplitude $a$. When the particle
activity falls on an ``even'' site the particle can make a move at once to
$y=O(+a)$ and can explore the whole chain in a time $O(N^{3/2})$ independent of
the value of $a$. Thus, there are configurations with arbitrarily high energy
with fast mixing.

For case (2) we consider the configuration
\begin{equation}
  y_i = N\sqrt{T/k} \cos{(2 \pi i/N)} \label{eq:cos}
\end{equation}

The algorithm advances by motion of particles a distance $O(\sqrt{T/k})$,
and the energy of this configuration is $O(N T)$.  On starting at an arbitrary
position the ECMC algorithm runs ``downhill'' to the smallest values of $y_i$,  Fig.~\ref{fig:cosfig},
and starts moving the particles with smallest $y_i$ in a positive direction,
``filling in'' the bottom of the cosine.  To
erase the initial configuration requires moving each of the $N$ particles a
distance $O(N \sqrt{T/k})$. Thus, we have a lower bound on mixing time of $N^2$,
when we require that the energy $O(N T)$ in the initial state, as was found for
the hard-rod problem.

\section*{Breaking Balance}
\begin{figure} 
\center{\includegraphics[width=.6\columnwidth]{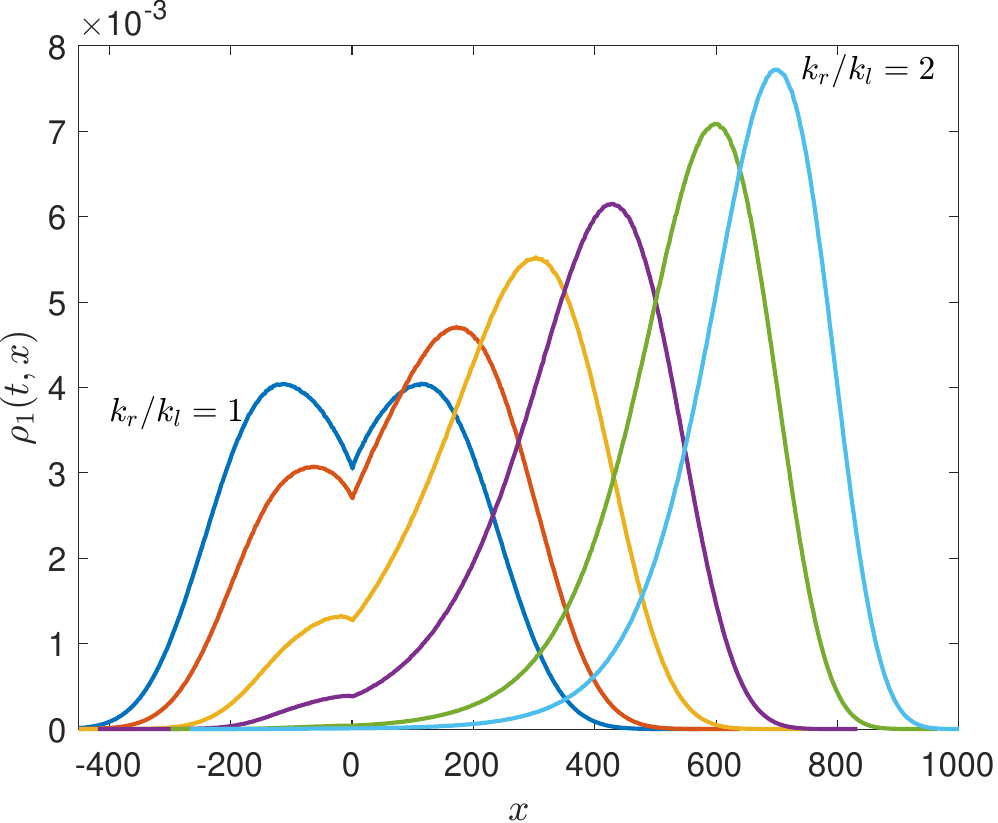}}
  \caption{Plots of $\rho_1(t,x)$ for a dynamical system that breaks global balance for ratios $ k_r/k_l= [1,
    1.1, 1.3, 1.5, 1.8, 2]$, $t=2048$, $T/k_l=1$.
    \protect\url{sargent:Balance/Balance/super2.m}
 }\label{fig:balance}
\end{figure}

Given the similarity of ECMC to active matter, with driven trajectories, it is interesting to modify
the dynamics in such a way that driven states are no-longer compatible with the
Boltzmann distribution. Are the notable features that we recognize for the true
self-avoiding walk still present in such a fully non-equilibrium systems. We choose
to do this by modifying the interactions in the harmonic chain in such a way that they
are non-longer reciprocal~\cite{fruchart}, that is when the force from $i$ to $i+1$, is
different from the force for $i+1$ to $i$. We modify the form
eq.~(\ref{eq:harmonic}) so that in the collision rules eq.~(\ref{eq:energy})
different constants $k_r$ and $k_l$ are used for interactions to the right and
to the left of the active particle. We perform simulations for several values of
$k_r/k_l$ and plot the results in Fig.~\ref{fig:balance}. It is not surprising that
distribution $\rho_1(t,x)$ drifts  in a manner similar to
Fig.~\ref{fig:l0}. However, remarkably the evolution again conserves a
singularity at $x=0$. We have no explanation as to the origin and the stability
of this singularity under several forms of perturbation of the true
self-avoiding walk. 

% In our previous study of factor fields we proposed a possible initial state
% that gives the longest time to reach equilibrium, thus defining a bound on the
% mixing time. For a set of impenetrable objects in one-dimension we proposed
% packing tightly all the particles, finding a mixing time in $N^2$. We can also
% ask ourselves a similar question for the Gaussian chain. Clearly if we inject
% an unbounded quantity of energy into the initial state we can produce states
% that are arbitrarily far from equilibrium. It seems more natural to find the
% most pessimistic state with an energy close to that of the equilibrium system
% $NT$. One possible starting state is a sinusoidal wave $y=a \sin{(2\pi i/N)}$
% which has an energy $~a^2/N$. If we put this equal to $NT$, we find that the
% amplitude should be $a\sim N \sqrt{T}$. Typical step size when applying ECMC
% $\sqrt{T}$. Filling argument then gives time to remove wave is $O(N^2)$ our
% estimate for the mixing time.

% breaking balance, weird forwards/backward rules.

% \section*{Other stuff}
% break balance, first-collision rule? noise in potential?  flip between
% potentials, back and forward potential different non-reciprocal interactions

% entropy from Haim~\cite{haim}, or energy

% skewed starting states and decay to equilibrium

% \begin{equation}
%   S(\vecq)=N^{-1}\langle |\trho(\vecq)|^2\rangle ,
%   \label{eq:S0}
% \end{equation}

% \begin{equation}
%   h[S(\vecq)] = \frac{1}{2{(2\pi)}^d\brho} \int d\vecq \left( \ln S - S
%     + 1 \right).
%   \label{eq:hS}
% \end{equation}

\section*{Conclusions}
We have studied variants of the ECMC including factor fields, in order to compare with an exactly
soluble model of polymer growth. We find that the distribution functions, which
were studied in detail in~\cite{nonrev} for a harmonic chain, remain valid even
for physical systems that are strongly heterogeneous, such as the Lennard-Jones
chain of Fig.~\ref{fig:config}. Introducing tension however leads to very
different distribution functions. It is remarkable however that the singularity
at the origin in the function $\rho_1(t,x)$, remains even as the average
position drifts. We have no explanation as to why this is so. Similar
singularities remain even when breaking global balance.

%\acknowledgments
\ack%iop
{I thank Werner Krauth for extensive discussions on 
 non-reversible simulations and lifted TASEP.}

\section*{Code}
C++ code for simulating the exponential and Lennard-Jones interactions,
available from \url{https://github.com/acmaggs/acmaggs.github.io}.

\bibliographystyle{ieeetr} %iop
\section*{References} %iop
\bibliography{avoid}
\end{document}